\documentclass[conference]{IEEEtran}
\IEEEoverridecommandlockouts

\usepackage{times}

\usepackage{amsmath,mathtools}
\usepackage{amssymb}

\usepackage{graphicx}
\usepackage{blkarray}
\usepackage[numbers]{natbib}
\usepackage{color} 
\usepackage{lineno}
\usepackage{url}
\usepackage{pdfpages}
\usepackage{hyperref}

\hypersetup{
    unicode=false,          
    pdftoolbar=true,        
    pdfmenubar=true,        
    pdffitwindow=false,     
    pdfstartview={FitH},    
    pdftitle={My title},    
    pdfauthor={Author},     
    pdfsubject={Subject},   
    pdfcreator={Creator},   
    pdfproducer={Producer}, 
    pdfkeywords={keywords}, 
    pdfnewwindow=true,      
    colorlinks=true,       
    linkcolor=blue,          
    citecolor=blue,        
    filecolor=magenta,      
    urlcolor=black           
}

\date{}

\begin{document}
	\title{Making an Example: Signalling Threat in the Evolution of Cooperation\\
		\thanks{This work was supported by the Future of Life Institute (grant RFP2-154).}
	}

	\author{\IEEEauthorblockN{Theodor Cimpeanu}
		\IEEEauthorblockA{\textit{School of Computing, Engineering \& Digital Technologies} \\
			\textit{Teesside University}\\
			Middlesbrough, UK \\
			T.Cimpeanu@tees.ac.uk}
	\and
	\IEEEauthorblockN{The Anh Han}
	\IEEEauthorblockA{\textit{School of Computing, Engineering \& Digital Technologies} \\
		\textit{Teesside University}\\
		Middlesbrough, UK \\
		T.Han@tees.ac.uk}}

\maketitle
\begin{abstract}  
Social punishment has been suggested as a key approach to ensuring  high levels of cooperation and norm compliance in one-shot (i.e. non-repeated) interactions. However, it has been shown that it only works when  punishment is highly cost-efficient. On the other hand, signalling retribution hearkens back to medieval sovereignty, insofar as the very word for gallows in French stems from the Latin word for \textit{power} and serves as a grim symbol of the ruthlessness of high justice. Here we introduce the mechanism of signalling an act of punishment and a special type of defector emerges, one who can recognise this signal and avoid punishment by way of fear. We describe the analytical conditions under which threat signalling can maintain high levels of cooperation. Moreover, we perform extensive agent-based simulations so as to confirm and expand our understanding of the external factors that influence the success of social punishment. We show that our suggested mechanism catalyses cooperation, even when signalling is costly or when punishment would be impractical. We observe the preventive nature of advertising retributive acts and we contend that the resulting social prosperity is a desirable outcome in the contexts of AI and multi-agent systems. To conclude, we argue that fear acts as an effective stimulus to pro-social behaviour.
\end{abstract}



%

\section{Introduction}
Punishment has been suggested as one of the most relevant explanations to understanding how selfish individuals self-organise and enforce cooperation or compliance to social norms in various societies \citep{key:fehr2002, Herrmann2008, powers2012punishment, Sigmund2001PNAS, boydPNAS2003}. Numerous empirical studies show human proclivity towards punishing unjust behaviour or violations of social norms, often at great cost to their own selves \citep{key:fehr2002, Herrmann2008, de2004neural}. Although in modern societies sanctioning systems have been widely implemented in the hopes of enforcing laws, many social norms continue to be upheld by the effects of private sanctions \citep{key:fehr2002}. Moreover, third-party punishment has also been implemented in various online systems, such as virtual agent societies \citep{savarimuthu2009social} or vendor marketplaces \citep{michalak2009exogenous}, as a method of enhancing pro-social behaviour and norms compliance, by both customers and sellers \citep{michalak2009exogenous}.

Cooperation often emerges under the influence of social punishment (i.e. punishing wrongdoers) \citep{cluttonbrock1995, key:fehr2002, Han2016AAAI}, but this fails to explain how punishment evolves, especially if it costly to punish others. Indeed, it has been concluded that punishment is often maladaptive within its respective games \citep{dreber2008, key:ohtsuki2009, WuPNAS2009} and that punishment can only evolve if it is cost-effective to do so (i.e. the offender suffers much more from retribution compared to the aggrieved party) \citep{dreber2008, key:ohtsuki2009, WuPNAS2009}. 

Refusing low offers in the ultimatum game (another form of punishment) in the presence of observers, made the wilful (punishers) more likely to receive higher offers in future interactions \citep{fehr2003}. The fear of having a low offer refused increased the tendency to present higher offers to obstinate individuals and this may help explain which mechanism(s) allow the promotion of punishment when it is costly to do so. In addition, pre-play signalling has been shown to open new avenues for cooperation to emerge, even when such signals are meaningless \citep{SANTOS201130}. One area, to our knowledge, which has not yet been explored is using signalling \citep{Huttegger10873} to explain the emergence of punishment as a viable strategy in evolutionary games, whereby punishers make their retributive deeds well-known in the population, as a deterrent to malefactors. Threat of punishment has also been indicated as one form of making credible commitments \citep{Nese2011-chapter,key:Hanetall_AAMAS2012}, which becomes another reasonable explanation to the dilemma of social punishment. 

In this paper, we propose and analyse a novel approach towards explaining the evolutionary advantage of punishers in the context of anonymous interactions \citep{ key:Sigmund_selfishness} (without relying on reputation). We make use of evolutionary game theoretic models \citep{Hofbauer1998, key:Sigmund_selfishness} (see Section \ref{models_methods}) to show that signalling acts of punishment can promote the emergence of cooperation in the selfish environment of the one-shot Prisoner's Dilemma (PD) \citep{key:Sigmund_selfishness}. This game is a popular underlying agent interaction framework for studying self-regarding agents and it is also the most difficult pairwise social dilemma for cooperation to emerge in \citep{Hofbauer1998}. Threat of punishment can reduce defection from others without having to punish and we show that social welfare in this regime is much higher than what traditional social punishment models suggest (see Section \ref{simulation_results}).  We provide a comprehensive view of the outcomes of external factors, such as cost of signalling or effectiveness of punishment, and we show that expensive signalling can still provide meaningful gains to cooperation when punishing others is costly.

The effect of threat of punishment by costly signalling may provide key insights into policy making in the context of distributed systems or artificial intelligence. Indeed, it has been concluded that increasing the probability of developing super-intelligent agents is incompatible with using safety methods that incur delays or limit performance \citep{bostrom2017}. What is more, when technological supremacy can be achieved in the short to medium term, the significant advantage gained from underestimating or even ignoring ethical and safety precautions could lead to serious negative consequences \citep{armstrong2015, cave2018}. One proposed solution to mitigating this dangerous behaviour is to look towards intrinsic measures of encouraging AI research communities to want to pursue safe, beneficial design methodology \citep{baum2016}. Our results show that threat signalling may serve as one intrinsic factor to prevent catastrophic consequences in that regard.

\section{Related Work  }
Punishment has been a major explanation for the evolution of cooperation in the context of the one-shot interaction \citep{key:fehr2002,de2004neural,powers2012punishment} (for other explanations, see a survey in  \cite{key:Sigmund_selfishness}). A critical condition for cooperation to be sustainable in evolutionary models \citep{Sigmund2001PNAS,boydPNAS2003,key:Hauert2007}, as well as observable in lab experiments, requires the punishment to be cost-efficient, i.e. the effect it has on the wrongdoer should be sufficiently large compared the cost issued towards the punisher.

Signalling within and between organisms has been investigated using game theoretic models in areas of biology, economics and philosophy and it has been suggested that certain qualitative aspects are common to many real-world interactions \citep{Huttegger10873}. Furthermore, it has been shown that signalling is a robust mechanism for promoting cooperative action in certain collective quorums \citep{pacheco2015plos}. In the presence of meaningless (no pre-defined meaning or behaviour) pre-play signals, cooperation has been shown to emerge as a result of individuals learning to discriminate between different signals and reacting accordingly \citep{SANTOS201130}. Pre-play signalling has also been studied in the context of the evolution of honest signalling, showing that honest signalling only emerges when signalling is costly \citep{Catteeuw2013}.
To the best of our knowledge, no work so far has studied how signalling theory could explain the prevalence of social punishment by advertising acts of punishment after the fact. 

Reputation has been suggested as an approach towards addressing this puzzle \citep{dosSantos2011, raihani2015reputation}, whereas agents' actions consolidate in the eyes of observers some assumption of future behaviour. In this manner, social punishers can benefit indirectly through maintaining a reputation of punishing unjust behaviour \citep{hilbe2012emergence}. However, the assumption that agents' actions are not anonymous proves unrealistic in many social contexts or application domains \citep{key:Sigmund_selfishness}, i.e. in very large societies or when observation is difficult. 

The simple presence of an audience has been show experimentally to increase human propensity for moralistic punishment, causing an increase in costly punishment as a response to perceived moral violations \citep{Kurzban2007AudienceEO}. Participants did not expect to encounter audience members again and the results hold for anonymous interactions or when the only observer was the experimenter. This suggests that there is at least some type of benefit to increasing the observability of one's willingness to punish, beyond reputation. Participants were generally not self-aware of the reasons for which they decided to punish, so in the context of self-organised societies \citep{boes2017}, this would explain why some individuals act towards the interests of society as a whole, irrespective of their intentions to do so \citep{Stewart_2017}. To this end, we are further motivated to study inherent normative mechanisms that have developed as a result of indirect evolutionary advantages. 

Survey data on contribution norms in homogenous and heterogeneous groups has demonstrated that uninvolved individuals hold well defined, yet conflicting normative views on equality, equity and efficiency \citep{reuben2013}. That being the case, it has also been shown experimentally that punishment can help groups overcome this collective action problem, through the emergence of strong and stable contribution norms \citep{reuben2013}. With regard to self-organised systems, punishment may help self-organising agents come to collective agreements on normative standards for efficiency, equity and equality. 

Finally, punishment and sanctioning have been studied extensively multi-agent system (MAS) literature \citep{villatoro2011dynamic, balke2012}). Differently from our work, these studies aim at using the cooperation enforcing power of the mechanism for the purpose of regulating individual and collective behaviour, formalizing different relevant aspects of these mechanisms (such as norms and conventions) in a MAS. Moreover, to the best of our knowledge, no work exists in the literature that analyses how sending costly threat of punishment can improve cooperation. As we show later, this mechanism can significantly enhance cooperation even when punishment is not highly cost efficient.

\section{Models and Methods} \label{models_methods}
We adapt the Prisoner's Dilemma (PD), first by integrating the option of costly punishment as a benchmark and we follow by describing the main model and the different configurations which we explore using replicator dynamics and simulations. By choosing the most competitive social dilemma \citep{Hofbauer1998}, we explore the toughest environment for the emergence of cooperation, therefore increasing the relevance of any observed effects.

\subsection{Prisoner's Dilemma (PD)}

The one-shot PD is characterised by the following payoff matrix: 
$$
\begin{blockarray}{ccc}
& \textsf{C} 	& \textsf{D} \\
\begin{block}{c(cc)}
\textsf{C}	& R 	& 	S  \\
\textsf{D}	& T 	& 	P  \\
\end{block}
\end{blockarray}.
$$
Players experience, in pairs, a cooperation dilemma. In an interaction, individuals can decide whether to cooperate (play C) or defect (play D). Mutual cooperation (mutual defection) yields the reward $R$ (penalty $P$), whereas unilateral defection provides a defector with the temptation $T$ and the cooperator with the sucker's payoff $S$ ($T > R > P > S$) \citep{key:Sigmund_selfishness}. The game is considered one-shot, in other words there is no memory of past actions or prior knowledge about the interaction. We note that an act of cooperation, i.e. \textit{playing} C is different from a player adopting the strategy C. For the latter, a player will always play C and this is likewise true for acts of defection and D players.

\subsection{Social Punishment without Threat}
We extend the PD by allowing a special type of C player the option of costly punishment, thereby becoming a punisher (P). After the normal interaction has taken place, a P player chooses to punish those opponents who played D during the interaction. A punishment act consists in paying a cost $p$ to make their opponents incur a penalty $q$. Contrary to previous work that  focuses mostly on efficient punishment \citep{Rand2010624, boydPNAS2003}, we include the case where $p > q$, in order to better understand whether and when highly costly or inefficient punishment can still act as a promoting mechanism of cooperation. The newly defined P strategy always cooperates with C (as well as other P) players and always punishes D players. By including this strategy, we can analyse the evolutionary dynamics of punishment strategies and their viability in the evolution of cooperation. The $3 \times 3$ payoff matrix for the strategies P, D and C (for row player), is given by:  
$$
\begin{blockarray}{cccc}
& \textsf{P} 		& \textsf{D} 		& \textsf{C} \\
\begin{block}{c(ccc)}
\textsf{P}	& R 		& 	S - p 	& R  \\
\textsf{D}	& T - q 	& 	P		& T  \\
\textsf{C}	& R			& 	S		& R  \\
\end{block}
\end{blockarray}
$$
\subsection{Social Punishment with Threat}

We extend the standard social punishment model by introducing signalling an act of punishment and responding to such signals. Firstly, we consider a new type of punisher (denoted by PT) who, upon punishing a defector, can advertise this act by paying a cost $\theta$, thereby alerting future opponents to their willingness to punish (and to the consequences of defecting against them). As such, a new type of defector arises (denoted by DT), who, once receiving the threat of punishment, will react and thus cooperate with the signalling punishers (to avoid punishment). PTs cooperate between each other, whereas DTs defect against each other, in similar fashion to P and D players. For infinite population size we can derive the $4 \times 4$ payoff matrix for PT, D, DT and C (for row player) as follows: 
$$
\begin{blockarray}{ccccc}
& \textsf{PT} 	& \textsf{D} 	& \textsf{DT} 				& \textsf{C} \\
\begin{block}{c(cccc)}
\textsf{PT}	& R 	& S - p - \theta 	& R 	& R \\
\textsf{D}	& T - q	& P					& P 	& T \\
\textsf{DT}	& R 	& P					& P 	& T \\
\textsf{C} 	& R		& S 				& S 	& R \\
\end{block}
\end{blockarray}
$$
In order to derive the payoff matrix for infinite populations,  notice that we can disregard the initial encounter between a PT player and either type of defector. Given some probability dependent on the composition of the population, the PT player can enact a punishment upon a DT player. We explain this interaction in-depth in Section \ref{simulations} and provide average payoffs in the case of finite populations (the above payoff matrix for  infinite populations can then be recovered at the limit of increasing the population size to infinity). As this population is infinitely large, the infinitesimally small initial interaction can be safely forgone for the sake of simplification.  

\subsection{Methods}

All the analysis and numerical results in this paper are obtained using evolutionary game theoretic methods, using replicator dynamics for infinite populations \citep{ Hofbauer1998} and agent-based simulation for finite populations \citep{key:Sigmund_selfishness,novaknature2004}. In this setting, the payoff for each agent represents their fitness or social success. Evolutionary dynamics are then shaped by social learning \citep{key:Sigmund_selfishness, Hofbauer1998}, whereby the most successful individuals tend to be imitated more often by others. 

\subsubsection{Replicator Dynamics} \label{replicator-dynamics}

Replicator dynamics are used to study the growth of each fraction (of strategies) in the population, as a function of their frequency and relative fitness, where the fitness in this case corresponds to their payoffs \citep{key:Sigmund_selfishness, Hofbauer1998}. If we consider a triple strategy game with PT , D and DT, we denote   
$x_{PT}$, $x_D$ and $x_{DT}$ the fraction of each strategy, respectively. Therefore, $x_{PT} + x_D + x_{DT} = 1$. The average payoff ($\Pi$) for each strategy reads:
\begin{equation}
\label{eq:rep_eq_3strats}
\begin{split}
\Pi_{PT} &= (1 - x_D)*R + x_D*(S - p - \theta), \\
\Pi_D &= x_{PT}*(T - q) + (1 - x_{PT})*P, \\
\Pi_{DT} &= x_{PT}*R + (1 - x_{PT})*P.
\end {split}
\end{equation}
In order to calculate the relative fitness, we require the average fitness ($\bar{\Pi}$) in the population:
\begin{equation}
\bar{\Pi} = x_P*\Pi_{PT} + x_D*\Pi_D + x_{DT}*\Pi_{DT}.
\end{equation}
We can then calculate the gradients of selection for each strategy with the replicator equations:
\begin{equation}
\begin{split}
\dot{x}_{PT} &= x_{PT}*(\Pi_{PT} - \bar{\Pi}), \\
\dot{x}_D &= x_D*(\Pi_D - \bar{\Pi}), \\
\dot{x}_{DT} &= x_{DT}*(\Pi_{DT} - \bar{\Pi}). 
\end{split}
\end{equation}
According to replicator dynamics, whenever a gradient is positive (i.e. $\dot{x} > 0$), the frequency of that particular strategy grows in the population. We can similarly extract the replicator dynamics for P, D, C and PT, D, DT, C. 

\subsubsection{Agent-Based Simulations} \label{simulations}

For our simulations, we adopt a population size $N = 100$ agents. At the beginning of the game, each agent is randomly assigned a strategy from all the available strategies for that experiment. We assume sequential interactions between players (ordered encounters), which is randomised at runtime. At each time step (generation), each agent plays the PD with every other agent in the population. The fitness for each agent is the sum of their payoffs from each interaction. 

Social learning is modelled using the pairwise comparison rule \citep{traulsen2006}, a standard approach in studying evolutionary dynamics in evolutionary game theory, which states that a player $A$ with fitness $f_A$ can imitate another player $B$ with fitness $f_B$ with a probability given by the Fermi function, i.e. $P_{A, B} = (1 + e^{-\beta(f_B-f_A)})^{-1}$, where $\beta$ represents the intensity of selection, i.e. how strongly the agents value the difference in fitness between them and their opponents. For $\beta = 0$, we obtain neutral drift (random decisions), whereas large $\beta$ values lead to increasingly deterministic imitation. We assume at most one imitation can happen per generation (asymmetric update).  

In the absence of exploration or mutations, evolution inevitably leads to monomorphic states. Once such a state has been reached, it cannot be escaped solely through imitation. Thus, we assume that, with a certain mutation probability $\mu$, an agent can randomly choose another strategy to adopt without necessarily imitating an existing agent.

We simulate the evolutionary process for $10^4$ generations and average our measurements over the final $10^{3}$ steps for a clear and fair comparison (for example due to cyclic patterns). Furthermore, the results for each combination of parameters are obtained from averaging 500 independent realisations, with the exception of typical run patterns. As we include mutations in this work, it is important to note that no simulation reaches a monomorphic state. 

As the scenario we are exploring involves a well-mixed population, we simplify the model using a statistical average of the payoffs for each conditional strategy, as opposed to carrying out the random ordering of interactions at the start of each game. The average payoffs for social punishment without threat remain the same as previously discussed in Section \ref{replicator-dynamics}. 

To derive the average payoffs, we consider two distinct sequences of events, for each agent acting out a conditional strategy (PT and DT). Firstly, we consider the case when one PT player encounters a D player at the start of the generation. In this instance, the PT will punish the D player, while also incurring the cost $\theta$ for signalling the act of punishment. Each subsequent interaction with DT players will result in a reward for both the players, as DTs will react to the signal and avoid defecting against that PT player. Conversely, we consider how the payoffs change when a PT player encounters a DT player as the first defection against that PT in that generation. As the PTs signal is unbeknownst to the DT, it will defect. In turn, the PT will carry out their act of punishment, causing both players to miss the opportunity of cooperating.  

The probability of either sequence happening first is dependent on the composition of the population at each first interaction for PTs. The payoffs for all other strategies remain unaffected. Let $n_1$, $n_2$ and $n_3$ denote the numbers of $PT$, $D$ and $DT$ players in the population, respectively.  We have $n_1 + n_2 + n_3 = N$. We denote $\Pi_{A,B}$ the payoff received by a player following the strategy A when facing players following strategy B (some payoffs are equivalent e.g. $\Pi_{C,C} = \Pi_{PT,C} = \Pi_{C,PT} = \Pi_{PT,PT} = R$). The average payoffs for PT, D and DT read:
{\small
\begin{equation*} 
\begin{split} 
\Pi_{{PT}} &= \!\begin{multlined}[t][6.6cm]
\frac{1}{N-1} \Bigl((n_1+n_3-1) * \Pi_{C,C} + n_2 * \Pi_{PT,D} + \\ 
\frac{n_3*(\Pi_{PT,D}-\Pi_{C,C})}{n_2+n_3}\Bigr), 
\end{multlined}\\
\Pi_D &= \frac{1}{N-1} \Bigl(n_1*\Pi_{D,PT}+(n_2+n_3-1)*\Pi_{D,D}\Bigr), \\
\Pi_{{DT}} &= \!\begin{multlined}[t][6.6cm]
\frac{1}{N-1} \Bigl(n_1*\Pi_{C,C}+\frac{n_3*(\Pi_{D,PT}-\Pi_{C,C})}{n_2+n_3} + \\
(n_2+n_3-1)*\Pi_{D,D}\Bigr).
\end{multlined}
\end{split}
\end{equation*}
}
Note that at the limit of infinite population size, $N \rightarrow \infty$, $x_P = \frac{n_1}{N}$, $x_D = \frac{n_2}{N}$, $x_{DT} = \frac{n_3}{N}$, we recover the equations in (\ref{eq:rep_eq_3strats}) for infinite population sizes.

For the four-strategy game (PT, D, DT, C), we introduce $n_4$ as the number of C players. Therefore, we have $n_1 + n_2 + n_3 + n_4 = N$. The payoffs for PT, D, DT and C then become: 
{\small
\begin{equation*} 
\begin{split} 
\Pi_{{PT}} &= \!\begin{multlined}[t][6.6cm]
\frac{1}{N-1}\Bigl((n_1+n_3+n_4-1)*\Pi_{(C,C)}+n_2*\Pi_{PT,D}+\\
\frac{n_3*(\Pi_{PT,D}-\Pi_{C,C})}{n_2+n_3}\Bigr),
\end{multlined} \\
\Pi_D &= \!\begin{multlined}[t][6.6cm]
\frac{1}{N-1} \Bigl(n_1*\Pi_{D,PT}+(n_2+n_3-1)*\Pi_{D,D}+
n_4*\Pi_{D,C}\Bigr),
\end{multlined} \\
\Pi_{{DT}} &= \!\begin{multlined}[t][6.6cm]
\frac{1}{N-1} \Bigl(n_1*\Pi_{C,C}+\frac{n_3*(\Pi_{D,PT}-\Pi_{C,C})}{n_2+n_3}+\\
(n_2+n_3-1)*\Pi_{D,D}+n_4*\Pi_{D,C}\Bigr), 
\end{multlined} \\
\Pi_{C} &= \frac{1}{N-1}
\Bigl((n_1+n_4-1)*\Pi_{C,C}+(n_2+n_3)*\Pi_{C,D}\Bigr).
\end{split}
\end{equation*}	
}
Note that the payoffs for unconditional strategies are never affected by the ordering of interactions. Analytically, the payoff for punishers who threaten depends on the number of defectors who respond to threatening signals in the population, specifically the ratio between the two different defecting strategies (i.e. $\frac{n_3}{n_2+n_3}$). We contend that this could remain true in practical scenarios. In other words, it is better for signalling punishers when future defectors can discriminate signals precisely, and this indirectly increases the payoff for sensitive defectors, as well. This would suggest that there is a synergistic relationship between signalling punishers and fearful defectors and, to an extent, neither would prevail without the other.

\section{Results} 
We study the potential of punishment and signalling strategies and their effects on evolutionary dynamics using three aforementioned scenarios: no threat (P, D, C), threat without cooperators (PT, D, DT) and threat with freely available strategies (PT, D, DT, C). 

\subsection{Replicator Dynamics for Infinite Populations}
\begin{figure}
	\centering
	\includegraphics[width=0.95\linewidth]{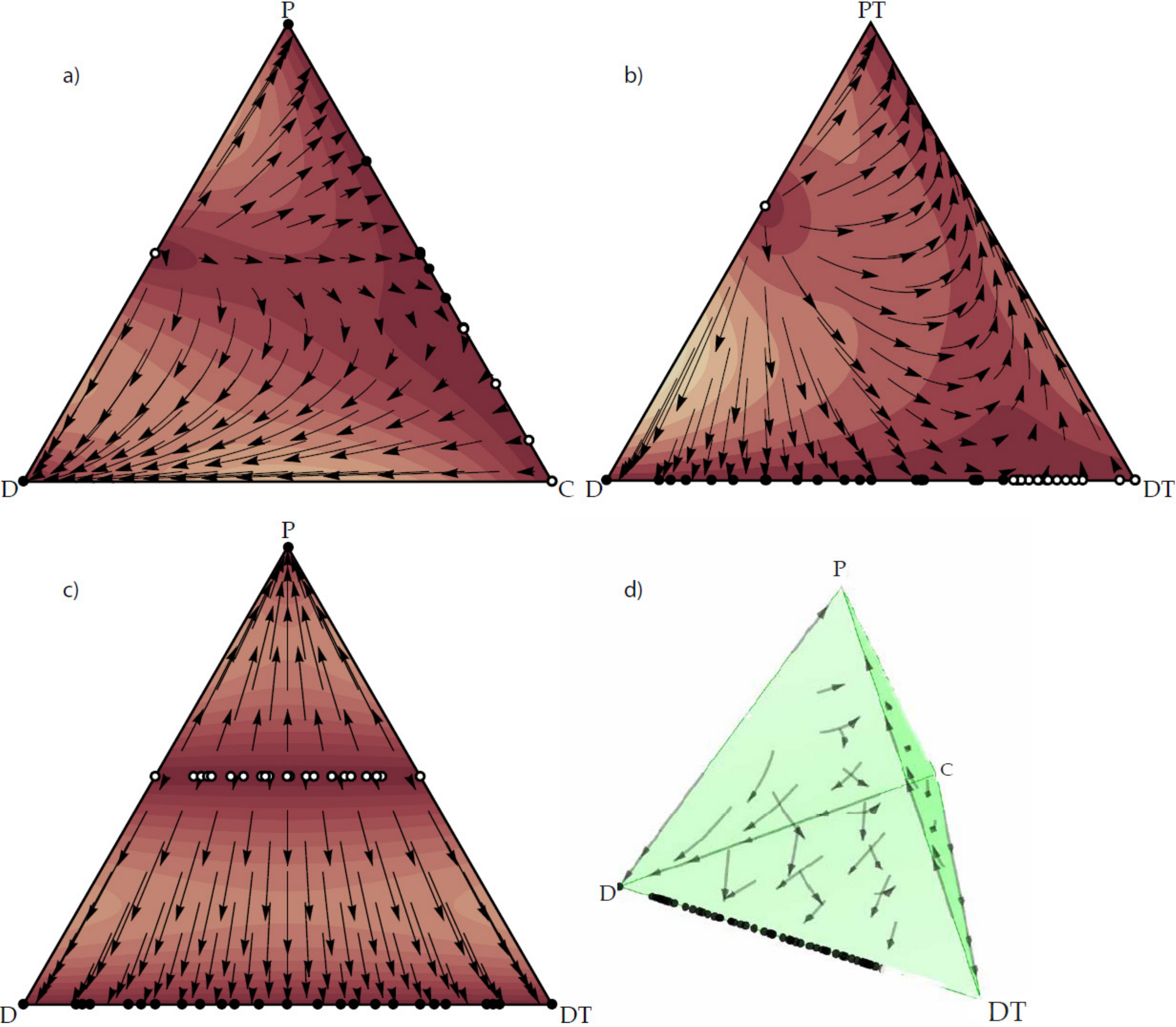}\vspace{-0.25cm}
	\caption{Phase diagram comparison between standard social punishment (left column) and social punishment with threat (right column) using replicator dynamics. Vertices represent specific strategies, whereas solid and empty dots stand for stable and unstable rest points, respectively. The colours represent speed of motion under the dynamic (lighter is faster, darker is slower). Parameters: $T = 2; \ R = 1; \ P = 0; \ S = -1; \ p = 1; \ q = 3; \ \theta = 1$.}
	\vspace{-0.5cm}
	\label{fig:dynamo}
\end{figure}
Following our analysis on infinite populations, we find that introducing threat signalling introduces a type of beneficial dynamic which fosters cooperation. The relationship between signalling punishers and defectors remains very similar to the one found in standard social punishment models (See Figure \ref{fig:dynamo}a and \ref{fig:dynamo}c). Increasing the efficiency ratio of punishment (i.e.  $\frac{q}{p}$) is the only way in which social punishment can remain relevant and this can often lead to undesirable consequences. On the other hand, the dynamic created between PTs and DTs naturally promotes cooperation. DTs lose out to PTs, as DT players do not cooperate amongst each other, but they do better than their defecting brethren, by reaping the rewards of their reluctance to defect against PTs. Even when the fraction of cooperators ($x_P + x_D$) becomes very low, the existence of DTs catalyses the conversion towards cooperation (see Figure \ref{fig:dynamo}b). By increasing the efficiency in the case of threat signalling, we found that the range of compositions which lead to all defectors is reduced even further. The ratio $q/p$ also favours DTs over Ds, which can provide another avenue towards cooperation that does not exist in the absence of signalling. 

In Figure \ref{fig:dynamo}d, we show that the results remain robust when we introduce C players, and in fact that the model is more resilient to compositions with high $x_C$ compared to traditional social punishment models, in which C outperforms P (when $p$ is high).

Analytically, we confirm the results from our replicator dynamics analysis by computing the rest points for each system of equations. For P, D, C, we found the following stable rest point on the PC edge which we can see mirrored in Figure \ref{fig:dynamo}a (at $x_P = \frac{1}{2}$):
{\small
\begin{equation*}
	\{x_{P} = \frac{P-S+p}{R-S-T+P+p+q}, \ x_D = 1 - x_P, \ x_C = 0\}.
\end{equation*}
}
\begin{figure}
	\centering
	\includegraphics[width=\linewidth]{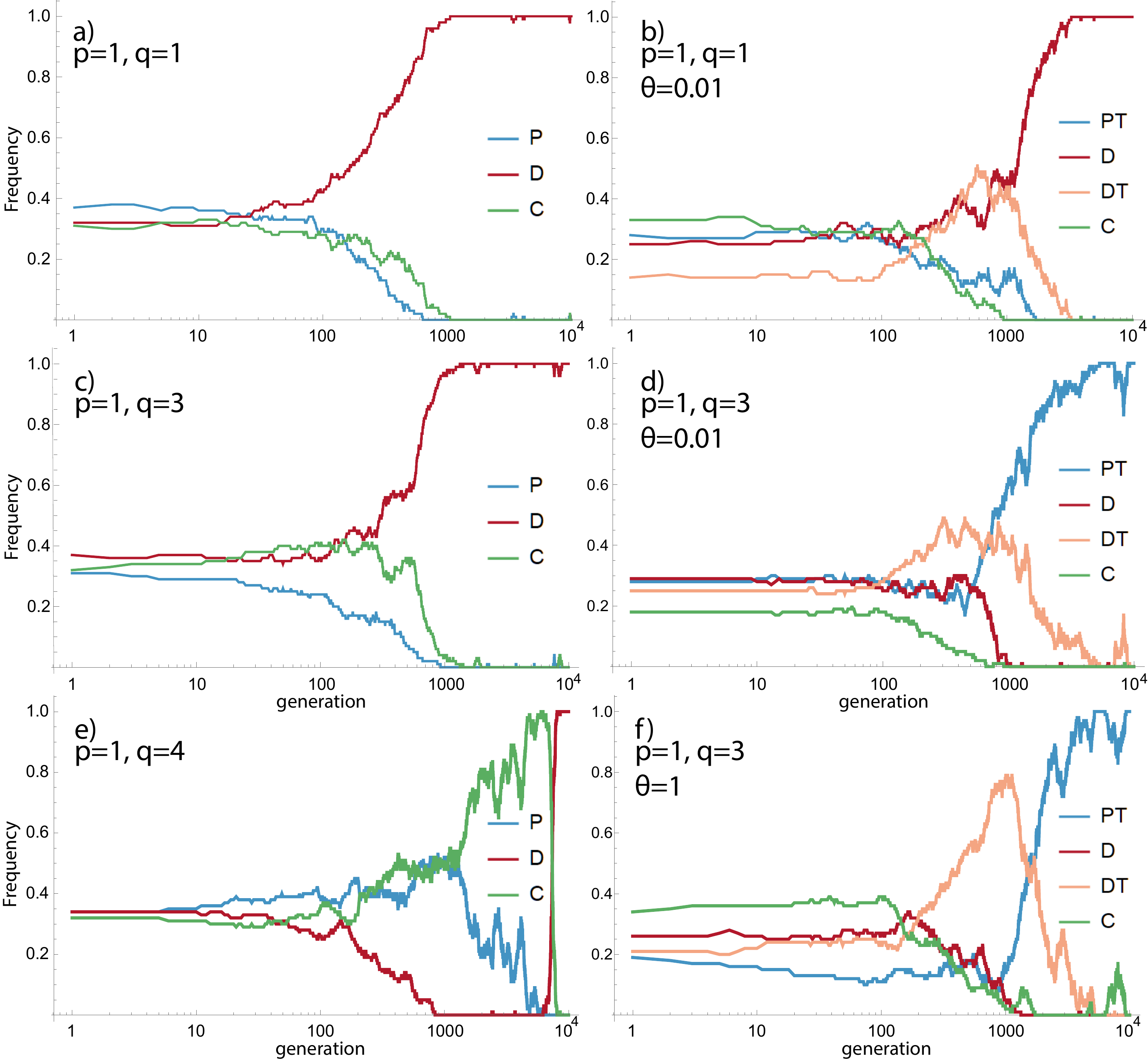}
	\caption{Typical evolution of frequency over time before (left column) and after (right column) introducing signalling of threat. Note that we only address the typical cases for clear comparisons. Parameters: $T = 2; \ R = 1; \ P = 0; \ S = -1; \ \mu = 0.001; \ \beta = 1$.}
	\label{fig:per-generation}
\end{figure}
Studying the case of PT, D, DT hints at the interesting dynamics seen in Figure \ref{fig:dynamo}b. We observe the following edge rest points: 
{\small
\begin{equation*} \label{restpoints-threat}
\begin{split}
	&\{x_{PT} = \!\begin{multlined}[t]
	\frac{P-S+p+\theta}{R-S-T+P+p+q+\theta}, \\ 
	x_D=\frac{R-T+q}{R-S-T+P+p+q+\theta}, x_{DT}=0 \} 
	\end{multlined} \\
	&\{x_{PT}=0, \ x_D=\frac{R-P}{R-S+p+\theta}, \ x_{DT}=\frac{P-S+p+\theta}{R-S+p+\theta}\} \\
	&\{x_{PT}=0, \ x_{DT} = 1-x_D\} 
\end{split}
\end{equation*}
}
We add that for the parameter values seen in Figure \ref{fig:dynamo}, we can observe a slight evolutionary advantage of PT against standard defectors, compared to traditional social punishers. For the same values of $p$ and $q$, while also having to pay a cost of signalling $\theta = p$, we observe the points $x_P = \frac{1}{2}$ without signalling and $x_{PT} = \frac{3}{5}$ after signalling. This shows that signalling social punishers have higher chances of survival against wrongdoers, even if the aggregated costs they expend towards punishment are larger than the ones paid by standard punishers. 
\begin{figure}
	\centering
	\includegraphics[width=\linewidth]{Figures/cooperation_aver_payoff_vary_p_q.pdf}
	\caption{Effect of the efficiency ratio of punishment on frequency of cooperation (punishers and cooperators alike) and average population payoff, before (left column) and after (right column) introducing inexpensive signalling. Parameters: $T = 2; \ R = 1; \ P = 0; \ S = -1; \ \mu = 0.001; \ \beta = 1; \ \theta = 0.01$.}
	\label{fig:average-coop-payoff}
\end{figure}
Note that we only discuss edge rest points for clear and concise presentation. Vertex rest points do exist in most scenarios, and they can be clearly seen in Figure \ref{fig:dynamo}. 

\subsection{Agent-Based Simulations for Finite Populations} \label{simulation_results}

The initial results from the agent-based simulations confirm the trend we found using replicator equations (see Figure \ref{fig:dynamo}). For typical runs, DT outperforms D which leads to the rapid extinction of D players and invites a booming growth of PTs (see Figure \ref{fig:per-generation}). When $q/p$ is low enough, we observed some exceptions to the norm, in which the initial conditions led to a population of predominantly DTs, which allowed PTs to flourish which, in turn, led to further defection. This type of cyclic behaviour only happened when punishing an act of defection and/or signalling it were extremely costly. We note that the results are robust even for costly signalling. 

When signalling is not costly, we found that cooperation is greatly increased across virtually all ranges of $p$ and $q$. In the case of simple social punishment, punishment is only effective at increasing cooperation when it is also efficient ($\frac{q}{p}>3$). On the other hand, even when punishment is very inefficient, we find a high propensity for pro-social behaviour in the presence of threat. Even when punishing is extremely costly ($\frac{q}{p}<1$), we observe close to 50\% cooperation. At moderate efficiency ($\frac{q}{p}>1.5$), the frequency of cooperation is very high ($\approx70\%$), comparable to highly efficient punishment without threat. 
\begin{figure}
	\centering
	\includegraphics[width=\linewidth]{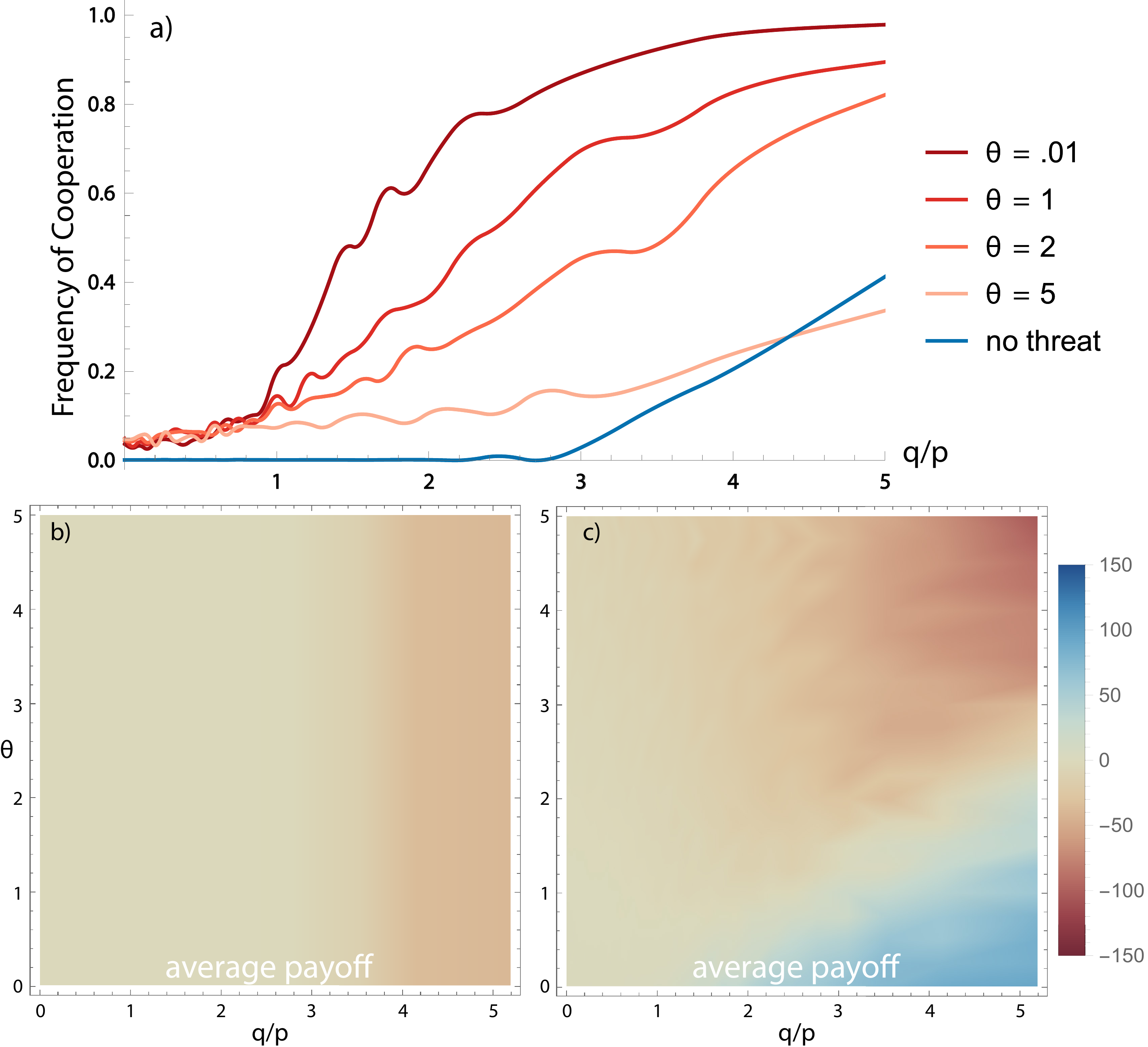}
	\caption{How costly signalling influences cooperation for (PT, D, DT, C) versus no threat (P, D, C). Figure a) shows the frequency of all cooperative strategies for various values of $\theta$ and for the standard case. Panels b) and c) show average payoffs as a function of efficiency of punishment and cost of signalling for standard no threat (left) and threat (right).  Parameters: $T = 2; \ R = 1; \ P = 0; \ S = -1; \ \mu = 0.001; \ \beta = 1; \ p = 1$.}
	\label{fig:cost-of-threat}
\end{figure}
We confirm the trends for higher values of $\theta$ (See Figures \ref{fig:per-generation}f and \ref{fig:cost-of-threat}a), but also observe one interesting outcome of signalling that may suggest a direct benefit towards the evolutionary promotion of punishing behaviour. Even at much higher ratios of efficiency of punishment ($\frac{q}{p}$), it is usually cooperators who prevail over the population, as seen in Figure \ref{fig:per-generation}e. Conversely, fearful defectors only cooperate with punishers, in the case of signalling, which naturally promotes the evolutionary advantage of punishers (see Figure \ref{fig:per-generation}f). One valuable side effect of this phenomenon is that the system is more resilient to mutation and exploration, a large population of punishers is more able to deal with defectors when compared to a large population of cooperators, where a single defector acts akin to a wolf among lambs. 

Welfare in the population (i.e. average population payoff) was overall much higher in our extended social punishment model (see Figures \ref{fig:average-coop-payoff}c, \ref{fig:average-coop-payoff}d, \ref{fig:cost-of-threat}b and \ref{fig:cost-of-threat}c). Of note, when punishment was very expensive ($p>2$), average payoff decreases dramatically in the extended model. Intuitively, this happens because PTs survive by cooperating with Cs and DTs, but also incur great costs in order to punish defectors and to deter DTs from following that trend. For higher values of $\frac{q}{p}$, we see another stark difference. With the exception of a very small region of values, welfare decreases ($\bar{\Pi}\approx-25$) as cooperation goes up, whereas the opposite is true for signalling ($\bar{\Pi}\approx75$). The \textit{power} difference between Ps and Ds proves damaging to social welfare - Ds do not cooperate, which causes Ps to lose fitness and in return, they pay a further cost, compounding the losses, causing Ds to incur even more loss. While this behaviour fosters cooperation, it greatly decreases social welfare. Inversely, a single act of punishment is enough to convert the entire population of DTs to cooperation, which is a qualifying factor in the growth of average payoff. 

Our comprehensive study of the external factors under which cooperation emerges, in regards to efficiency of punishment and the cost of signalling shows that fear of punishment enhances cooperation for almost all configurations (with the notable exception of highly efficient punishment coupled with expensive signalling). The results suggest that the transparency of social punishment, specifically the awareness agents have regarding acts of retribution, coupled with the ease of advertising said acts, behaves as a fulcrum towards cooperation. Fear of punishment, therefore, is most effective when awareness of who is or is not a punisher is high. On the other hand, the more deleterious an act of punishment is, the more likely it becomes for standard costly punishment to lead towards satisfactory outcomes. 

We also show that social welfare increases when signalling is not very costly, irrespective of the punishment efficiency. Interestingly, at higher $\theta$ values, social welfare lowers even as cooperation increases. We show that when punishment is effective, signalling can lead to lower levels of welfare. Intuitively, this suggests that advertising an ineffective act of punishment is productive even when signalling is expensive.

\section{Conclusions}
Punishment used as a deterring mechanism to prevent further damaging actions against the punisher or their peers appears to be a commonly found behaviour in human society and even in some animal hierarchies \citep{cluttonbrock1995}. Much of recent literature has concluded, however, that punishment may have evolved for reasons other than the promotion of cooperation, because significant benefits to punishers could typically not be found in the context of game theory \citep{ dreber2008, Rankin2009}. Indeed, it may be the case that even if punishing defectors incurs an immediate cost, it discourages observers from repeating said action, as long as the accumulated costs of punishment are outweighed by the additional acts of cooperation evoked over long runs \citep{dosSantos2011}. Our models suggest that this does not only happen in repeated interactions and that punishment can evolve through advertising the acts of punishment.

We show that signalling acts as a catalyst for the emergence of cooperation when defectors are fearful of the punishers who advertise themselves as such. Furthermore, we argue that exhibiting deeds of punishment can explain the success of punishers, when traditional social punishment mechanisms would otherwise fail due to external factors, such as lowly efficient acts of punishment. Indeed, it seems to be the case that fearing punishment can discourage future defectors even more than the evolutionary dynamics associated with inexpensive, deleterious deeds of retribution. Moreover, we show how the traditionally damaging effects of social punishment upon social welfare can be mitigated by way of threat. Because signalling punishers cooperate indiscriminately, they outperform fearful defectors who are always vying for higher status at the expense of others, including themselves. 

The prosperity of the population observed under threat of punishment speaks for the preventive nature of advertising acts of justice. Undeniably, it is a beneficial outcome for wicked ventures not to occur in the first place, but contexts such as the development of AI or climate change provide us with unparalleled incentive to prevent potentially disastrous consequences. Given the importance of intrinsic factors that guide the decisions of researchers and policy makers in the field \citep{baum2016}, we aim to explore further how the concept of threat, and the self-preservation associated with it, could help guide the current literature on this issue. Additionally, implementing this type of signal response could improve safety conditions in MA systems such as artificial societies \citep{balke2012, villatoro2011dynamic}, especially in cases where the transparency of interactions is reduced. 

These observations raise important questions around the coexistence of various types of punishers with different proclivities to signalling, as well as thresholds under which they decide that advertising their deeds of punishment would be appropriate. Reciprocally, defectors could evolve to decide when avoiding punishment is worth the act of justice and which punishers they can exploit even as they signal their propensity towards justice. Perhaps having a loud voice would be more conducive to the emergence of cooperation than the ease with which one can act revenge upon their enemies. Analytically, we have suggested the synergistic mutuality between signalling punishers and fearful defectors. We reason that the high sensitivity of defectors to signals of threat may allow less expensive signalling, whereas lowly responsive defectors require more obvious (and inherently costly) displays. 

Future work on this topic will aim to discover the complicated relationship between punishers with different inclinations towards signalling, as well as investigate defectors who judge which punishers are worth appeasing, based on the force of their retribution and the \textit{volume} of their signals. 
We are also interested in how signalling  of punishment enables  more cost-efficient cooperation-enhancing interference mechanisms \citep{han2018cost,cimpeanu2019exogenous}, reducing  costly incentivisation. Assuredly, it would contribute to the breadth of this research to study how network dynamics influence these observations and whether this model for social punishment is robust across a wide range of network topologies and population structures. 

\bibliographystyle{plain}  
\bibliography{bibliography}  
\end{document}